\documentclass[preprint]{aastex}  
  
\newcommand{\arcs}{\hbox{$^{\prime\prime}$}}  
\shorttitle {Infrared Starforming Disk in He2-10}  
\shortauthors{Beck, Turner\& Gorjian}

\begin{document}  
  
\title{Infrared Emission from Clusters in the Starforming Disk of He2-10}  
  
\author{Sara C. Beck\altaffilmark{1}}  
\affil{Department of Physics and Astronomy\\ Tel Aviv University, Ramat Aviv,   
Israel}  
\email{sara@wise1.tau.ac.il}  
  
\author{Jean L.Turner\altaffilmark{2}}  
\affil{Department of Physics and Astronomy\\ UCLA, Los Angeles, CA 90095-1562}  
\email{turner@astro.ucla.edu}

\and  
\author{Varoujan Gorjian\altaffilmark{3}}  
\affil{JPL, Caltech, MS 169-327, 4800 Oak Grove, Pasadena, Ca. 91109}  
\email{vg@coma.jpl.nasa.gov}  
  
\begin{abstract}  
We have made subarcsecond-resolution images of
 the central 10\arcs\ of the Wolf-Rayet dwarf galaxy He 2-10 at
11.7 microns, using the Long Wavelength Spectrometer on the Keck Telescope. 
The spatial distribution of the infrared   
emission roughly agrees with that of the rising spectrum radio sources seen by   
\citet{ko99} and confirms that those sources are compact HII   
regions rather than SNR or other objects. The infrared sources are
more extended than the subarcsecond rising spectrum radio sources, although the 
entire complex is still less than 5\arcs\ in extent. On sizescales of 
1\arcs\ the infrared and radio emission are in excellent agreement,
with each source requiring several hundred to a   
thousand O stars for excitation. The nebulae lie in a flattened   
disk-like distribution about 240 by 100 pc and provide all of the flux   
measured by IRAS for the entire galaxy in the 12 micron band; 30\% of the  
total IRAS flux from the galaxy emanates from one 15-30 pc source.  In this galaxy, 
intense star formation,   
probably triggered by an accretion event, is confined to a central disk which   
breaks up into distinct nebulae which presumably mark the sites of
 young super star clusters.
\end{abstract}  
  
\keywords{galaxies: individual (He 2-10) --- galaxies: starburst ---  
galaxies: star clusters --- galaxies: dwarf --- galaxies: peculiar  
--- radio continuum: galaxies--infrared}  
  
\section{Introduction}  
Henize 2-10 is a dwarf elliptical galaxy with a starburst nucleus and the first   
Wolf-Rayet galaxy identified.  Gas motions in and around He 2-10   
suggest that it is in an advanced stage of accretion,  which may be the trigger   
for the starburst   
\citep{ko95}.   He 2-10 differs from most dwarf starburst galaxies in that it has at   
least solar \citep{co94} to twice solar  
\citep*{be97} metal abundances and is relatively rich in molecular gas. Single-dish radio   
continuum fluxes of the galaxy as a whole have the non-thermal spectral index of   
-0.56 \citep{ko99}, 
similar to spiral galaxies and much steeper than most dwarf starbursts.  
Its distance is determined only to be between 6   
and 14 Mpc; we will use 9 Mpc here.     
  
The starburst in He 2-10 takes the form of a disk     
5-8\arcs\ in length and 2-3\arcs\ in height.  \citet{co94} found optical and UV   
super-star clusters lying in this disk, and \citet{be97}  
showed that the near-infrared J, H, and K emission traces it.   The disk is  
perpendicular to the bi-polar outflow of ionized gas described in \citet{me99}   
and to the apparent angle of the in-falling molecular gas \citep{ko95, me01}.   
\citet{ko99} measured the radio continuum at 6 and 2 cm with sub-arcsec   
resolution and found the same disk-like structure breaks up into 6 sub-sources,   
5 of which are strong enough to be fit and analysed individually, and all of   
which are stronger at 2 cm than 6 cm. A rising spectrum can be the sign of a   
radio source that is at least partly optically thick at the longer wavelength.   
  
\citet{ko99} concluded that the rising spectrum sources are probably giant HII   
regions excited by hundreds of O stars and containing such dense gas that their   
intrinsically thermal $(S_{\nu}=\nu^{-0.1})$ spectrum has been distorted by   
optical depth effects. Such sources have been seen in NGC 5253 \citep*{tu98, tu00} and   
NGC 4214 \citep{be01}. Near-flat or rising radio spectra are also seen, however, in   
non-thermal sources not excited by young stars, including AGN and SNR.  The best   
test for the presence of young stars is the infrared flux:  HII regions will   
have strong and predictable IR emission and the other   
candidates will not \citep{go01}. It is a particularly good way to weed out
SNR, which are weak IR emitters but strong radio emitters, and often present
in large numbers in starburst regions. The infrared is also the wavelength of choice   
for probing star formation regions in general, and can be expected to be   
particularly useful in He 2-10 where the radio and optical appearance, although   
similar in outline, are greatly different in detail, and where the optical   
extinction is known to be very high \citep{be97}. \citet{sa97} imaged He 2-10
in the mid-IR with 1.1" resolution at the CFHT; their raw images at 11.7 $\mu$m
show two main IR sources possibly associated with 4 of the 5 images; deconvolution
reveals a possible third,
weaker source to the east of the main emission. To better match the resolution
of the radio observations, we have accordingly obtained   
high resolution ($\sim$0.3\arcs ) middle-infrared   
(11.7 micron) images of the central disk in He 2-10 using the Long Wavelength   
Spectrometer \citep[LWS;][]{jb93} on the Keck Telescope. We describe the observations   
and results in the next section and discuss the peculiarities of He 2-10 in   
section 3.  
  
\section{Observations and Results}  
\subsection{The Infrared Observations}  
He 2-10 was observed on 17 February 2000 at the Keck I Telescope, using the   
LWS in imaging mode with a 1 micron wide filter centered at 11.7 microns.  The   
standard stars $\beta$ Gemini and $\alpha$ Bootes were used for calibration. The 
sky was unstable and non-photometric; the 
internal scatter in the standards was 7\% or lower over time spans of an hour 
but showed jumps of 18\% over longer periods.  
The calibration is further complicated by characteristics of the chip.
The pixel size is 0.08\arcs\ and the chip is 128 x 128, and the resulting 
10\arcs\ field is close enough to the size of the source that it is very difficult 
to find and subtract the sky background, a problem complicated by fringing at 
the edge of the 
chip and significant pixel-to-pixel variations in the noise. For these reasons 
we conservatively estimate the error in the total fluxes assigned to each source to be 15-20\%.
The angular resolution, 
based on the FWHM of the standard stars, is 0.3\arcs\ to 0.5\arcs.   
Two exposures of 216 on-source seconds each were obtained of He 2-10; they were   
added with no smoothing and the resulting image, rotated to   
the conventional NUEL orientation, is shown in Figure 1. In Figure 2 the infrared 
image is shown with the 2 cm radio contours (kindly provided 
by H.A. Kobulnicky) 
superimposed. Since the absolute registration of the infrared image is good only 
to about 1\arcs\ the 
superposition was adjusted by eye. 
  
\subsection{The Infrared Sources: Luminous and Dense HII Regions in a Disk} 
The infrared image of He 2-10 shows 3 sources in a roughly east-west line. The 
eastern and western infrared sources are individually extended east-west also,
following the  5 smaller radio knots 
seen by \citet{ko99}. We will therefore treat the infrared   
sources as emitted by 3 sources, which from east to west we call A, including   
radio knots 4 and 5, B which   
corresponds to radio knot 3, and C which includes radio knots 1 and 2.  In Table   
1 we give the infrared flux   
of each extended source, along with the radio flux for the region measured from
the 2 cm map for the extended infrared region, and the ratio of radio to infrared   
emission. Our 2 cm fluxes are those of the entire extended source and 
therefore larger than those of \citet{ko99}, which were obtained by fitting Gaussian
sources to the central cores of the nebulae. The mid-infrared fluxes we obtain 
are in good agreement with those observed at lower resolution by \citet{sa97}, 
although we find that the infrared sources are more extended than predicted 
by their deconvolution routines, and slightly higher than the values found
by \citet*{te93}.
   
The ratio of infrared to thermal radio flux from an HII region depends on   
the type and temperature of the dust. Lyman $\alpha$ dust 
heating alone gives a ratio of 10, but the value usually observed  
in starburst galaxies and Galactic star formation regions is between 
120-250, the excess generally due to dust heating by longer wavelengths and
optical depth effects in the radio. 
>From Table 1 we  
see that all the sources in He 2-10, with values $\sim$80,
are definitely in the regime of HII regions and not, for example, 
SNR whose infrared fluxes are many orders of magnitude weaker. This   
confirms the identification of the radio sources as compact HII regions 
\citep{ko99}. The relative fluxes of all three components are identical in 
the radio and infrared: 67\% of the flux is in component A, 8\% in B, and
22\% in C. 

While the infrared and radio emission in the three components agree over sizescales of 
$\sim1$\arcs\, 
their distribution is different in detail.  The radio emission is more compact than is the infrared. 
The radio sources have cores of $\sim0.1-0.5$\arcs\ \citep{ko99}, surrounded by flatter extended halos 
so the total sizes are similar to the infrared. The infrared sources are larger; when deconvolved from the $0.3--0.5\arcs$ PSF, 
component A is $\sim0.9\times0.5$\arcs\
or $40\times20$pc, B is $\sim0.3$\arcs\ or 13 pc, and C is $\sim0.5$\arcs\ (22 pc).  
The radio beam is about twice as large
as the infrared NS and slightly larger EW, yet the radio sources appear more compact, and the infrared sources do not
have the core and background structure of the radio.
  
A remarkable result from the infrared image is that the   
clumpy disk appears within the uncertainties to provide least 80\% of 
the total flux seen by IRAS at   
12 $\mu$m with a beam that included the entire galaxy. This total is that of   
the 5 clumps together; we do not see an extended smooth infrared component.   
He 2-10 is another case    
(like NGC 5253, \citet{tu00}) of a starburst dwarf in the great majority of
OB star   
formation is concentrated in a small volume, but unlike NGC 5253 the star   
formation activity is not in a single super cluster but in a much more complex   
geometry. If we fit a modified blackbody spectrum with $n \sim 1.5$
to our 11.7 $\mu$m and IRAS fluxes, the observed 11.7, 12, and 25 $\mu$m fluxes
can be fit with a 120K source. This 120K source would account for most of the IRAS
12 and 25 $\mu$m flux and thus 40\% of the total IRAS luminosity of
$\rm L_{IR} = 6.2 \times 10^9~L_\odot$.  

Using a reference OB star luminosity of $2.5\times10^5 L_\odot$ for an O7   
star, we calculate the OB star content of each infrared source from its
mid-infrared luminosity as listed in Table 1. We also list the OB star content
derived from the radio fluxes, using the assumption that they are optically
thin \citep{tu94}. If the HII regions have optically thick components, or if dust absorbs
a significant number of the photons, then this will be an underestimate of
the number of massive young stars. Note that   
the numbers of OB stars found for the three components based on these two
very different methods for the radio and infrared agree very well. 
The mid-infrared luminosities of regions A, B, and C agree with   
the ionization requirements of the radio knots, which argues that the radio
emission cannot be very optically thick on these sizescales. The rising radio spectra 
must be attributed only to the central cores of the nebulae.    
  
The radio/infrared sources contain very dense gas, as \citet{ko99} calculated   
from the radio optical depths,    
and a great deal of dust to account for their high extinction and infrared flux,   
and must therefore be very young. If one estimates a dynamical age
based on the time it would take them to expand to the stage of having a normal 
radio spectrum, one obtains ages of 500,000 years    
\citep{ko99}. However this assumes that there are no other forces working against their   
expansion. The apparent contradiction between the very short dynamical lifetimes   
of compact HII regions and their ubiquity has been reviewed by   
\citet{ku00}. The possible confinement mechanisms suggested for NGC 5253 
by \citet{tu00, go01}(i.e. magnetic fields of the milligauss level, gravity)   
may also be at work in He 2-10.  In the case of He 2-10, the more traditional
suggestion of confinement by dense molecular gas is also a viable
source of extra overpressure which can keep the   
nebulae confined and the gas density high.    
  
The age of the He 2-10 nebulae can be estimated independently from the   
mid-infrared line spectrum of \citet{be97} and the ionization of a   
starburst as modelled by \citet{cr99}. The models can reproduce the infrared   
line spectrum if   
the starburst in He 2-10 is relatively old, from 0.5 to $2\times10^7$   
years \citep{be97}.  This is not only much older than the dynamical age, 
a problem which   
appears in many other sources, it is as old or older than the super star   
clusters observed in the optical and UV in the same galaxy \citep{jo00}.  It is   
very hard to find a physically sensible scenario in which the optically visible   
star clusters, which have already dispersed their gas and dust, are younger than   
the infrared and radio sources which are still embedded in gas and dust.  It is   
possible that the age derived from the infrared spectrum is misleading. First,   
the infrared spectrum may be softened by higher than solar metallicity so the   
clumps may  
contain younger and hotter stars than the rather cool ones found by the models.    
Second, the ages found from these models assume that the burst formed with very   
massive stars present and that the soft ionization currently observed means that   
stars more massive than 35$M_\odot$ have disappeared.  The models cannot at   
present distinguish between a burst of $5\times10^6$ to $2\times10^7$ years with   
an initial upper mass cutoff of   
60$M_\odot$ and a much younger burst whose Initial Mass Function did not extend   
past 35$M_\odot$ to begin with.  Middle-infrared diagnostics of stellar age are   
greatly needed, as only in the middle-infrared can these very young and obscured   
HII regions be observed properly. We can only  
conclude that while these compact nebulae are doubtless the youngest part   
of the starburst their age cannot be determined to better   
than a factor of 10.   
  
\subsection{Accretion-Induced Star Formation Throughout a Dwarf Galaxy Disk?}  
The high-resolution radio, infrared and optical observations of He 2-10 find an   
extended disk that contains or is made out of many clumps, probably star   
clusters, some of which can be seen in the optical and the others only at the   
longer wavelengths. Other dwarf galaxies are known   
to contain  intense and small infrared and radio clumps excited by the youngest   
star clusters but they have either only one young, embedded cluster  
(NGC 5253)  or a few clusters with no clear underlying structure (NGC 4214,   
\citet{be01}).  He 2-10 is   
so far unique among dwarf galaxies in that its youngest star clusters are   
associated with a disk. The cluster distribution in He 2-10 in in fact   
reminiscent of that seen in spiral starburst galaxies like NGC 253.  Why is He   
2-10 so different from other galaxies of its apparent class?  We note that the 
extent of the ``disk'' (or linear structure) is $\sim$ 200 pc, possibly more if there is signifiant extension in the line of sight, which corresponds
a sound-crossing time of at least  $\sim 2 \times 10^8$ years. This is far
greater than the inferred ages of the compact HII regions and suggests a 
star formation trigger.  The infrared structure may reflect the rotating molecular gas disk whose morphology and kinematics are described by \citet{ko95}.  If molecular gas accretes onto a galaxy with a rotating disk, it   
will not matter where or at what angle it  
starts, it quickly ends up falling onto the disk.  So the accretion induced   
star formation will be concentrated in the disk, where local conditions and
perhaps self-gravity structures will   
determine the size and nature of the star clusters that will form.     
  
Another feature of He 2-10 which differs from the usual dwarf starburst galaxy   
is its metal content, which is around twice solar (from the NeII flux observed   
by \citet{be97}),   
rather than the approximately  0.1 solar seen in NGC 5253, II Zw 40 and many   
others. This may have an effect on the future evolution of the observed infared   
star clusters.  Young stars   
have violent mass outflow episodes which can partly disrupt a cluster and these   
winds are stronger at higher metallicity.  We speculate that the clusters in the   
disk of He 2-10 will in a few tens or hundreds of Myr be too small and faint to   
be true globular clusters.  Even without the effect of winds it is hard to tell what kind of   
clusters  
these sources will become, because of the relatively low spatial resolution we   
have at the distance  
of He 2-10.  Source A, for example,   
contains about twice as many OB stars as does the supercluster in NGC 5253   
\citep{tu00}, more than enough stars  
to evolve into a globular cluster, but we don't yet know if it contains more   
sub-sources than  
the two already seen and what the volumes and stellar densities will be.   
  
\section{Conclusions}  
Mid-infrared images of the dwarf starburst galaxy He 2-10 were obtained with the   
LWS on the Keck Telescope.  The images show infrared clumps along the galaxy   
disk which coincide with the rising spectrum radio sources seen by \citet{ko99}.   
The image shows that the mid-infrared emission is more extended than the 
radio emission, but is otherwise in excellent agreement. This confirms that 
the sources are compact HII regions excited by large and extremely young 
super star clusters. 
At least 80\% of the middle-infrared flux of the galaxy, as seen in IRAS, comes from the    
vicinity of these compact nebulae which extend over 200$\times$100 pc, and which emit $\sim$40\%
of the total infrared luminosity of the galaxy.  He 2-10 is another example of   
a starburst galaxy where the current star formation activity is very intense and   
confined to a very small volume.  We suggest that the disk is the site of the   
starburst because that is where the accreted molecular gas is falling, triggering
the burst. We also suggest that due to the high metal content
the clusters are unlikely to survive the 10 Gyr to become true globulars.   
  
\acknowledgments  
This work was supported in part by NSF grant AST-00 71 276 and a COR grant from   
the UCLA Academic Senate to J.L.T. and  
by the Israel Academy Center for Multi-Wavelength Astronomy grant to S.C.B. We thank H.A. Kobulnicky for the 2 cm radio map.

\clearpage   
  
%% No more than seven \figcaption commands are allowed per page,  
%% so if you have more than seven captions, insert a \clearpage  
%% after every seventh one.  
  
%% There must be a \figcaption command for each legend. Key the text of the  
%% legend and the optional \label in curly braces. If you wish, you may  
%% include the name of the corresponding figure file in square brackets.  
%% The label is for identification purposes only. It will not insert the  
%% figures themselves into the document.  
%% If you want to include your art in the paper, use \plotone.  
%% Refer to the on-line documentation for details.  

\figcaption{The LWS image of He 2-10 on the left, and the same image overlaid with the 2 cm radio contours on the right.   
The contour interval is 0.2 mJy/bm and the levels $\pm2^{n/2}$. \label{fig1}}

%% Tables should be submitted one per page, so put a \clearpage before  
%% each one.  
  
%% Two options are available to the author for producing tables:  the  
%% deluxetable environment provided by the AASTeX package or the LaTeX  
%% table environment.  Use of deluxetable is preferred.  
%%  
  
%% Three table samples follow, two marked up in the deluxetable environment,  
%% one marked up as a LaTeX table.  

%% In this first example, note that the \tabletypesize{}  
%% command has been used to reduce the font size of the table.  
%% Note also that the \label command needs to be placed   
%% inside the \tablecaption.  
  
\clearpage  
  
\begin{deluxetable}{crrrrrrr}  
\tabletypesize{\scriptsize}  
\tablecaption{Observed and Derived Cluster Parameters \label{tbl-1}}  
\tablewidth{0pt}  
\tablehead{  
\colhead{Infrared Source} & \colhead{A}   & \colhead{B}   
  &  
\colhead{C}  
  
}  
\startdata  
Radio equivalent$^a$ & 4 \& 5 & 3 & 1 \& 2 \\  
11.7 $\mu$m flux (Jy) & 0.59 & 0.07 & 0.22 \\  
2 cm flux (mJy)$^a$ & 7.27 & 0.89 & 2.76 \\   
IR/2 cm & 81 & 77 & 80 \\
$\rm L_{mIR}^b$ (120 K) & $1.7 \times 10^9$ & $2.1 \times 10^8$ & $6.5 \times 10^8$ \\
$\rm L_{IR}^c$ (IRAS) & $4.1 \times 10^9$ & $5.0 \times 10^8$ & $1.4 \times 10^9$ \\
O7 stars (radio)$^d$ & 5300 & 650 & 2000 \\
O7 stars (IR) $^d$ & 8500 & 1000 & 3300 \\

 \enddata  
  
%% Text for table notes should follow after the \enddata but before  
%% the \end{deluxetable}. Make sure there is at least one \tablenotemark  
%% in the table for each \tablenotetext.  
  
\tablenotetext{a}{\citet{ko99}. 2 cm fluxes are measured for the extended
regions corresponding to the infrared sources.} 
\tablenotetext{b}{For the 120K component fitting the 11.7$\mu$m and IRAS
12 and 25 $\mu$m fluxes, as discussed in the text.} 
\tablenotetext{c}{On the assumption that the FIR is distributed as the 11   
micron, discussed in the text.}  
\tablenotetext{d}{Using $\rm N_{Lyc}=9\times10^9(D/Mpc)^2~(S_{2\,cm}/mJy)$
\citep{tu00}; $L_{IR}/N_{Lyc}=2\times 10^{-44}$ for mid-IR (120 K component) 
estimate, and 1 O7
star = $\rm 10^{49}~s^{-1}$.} 
\end{deluxetable}

\end{document}